\documentstyle[12pt,aasms4]{article}
\begin{document}
\title{The Long and the Short of Gamma-Ray Bursts}
\author{J. I. Katz and L. M. Canel}
\affil{Department of Physics and McDonnell Center for the Space Sciences}
\authoraddr{Washington University, St. Louis, Mo. 63130}
\authoremail{katz@wuphys.wustl.edu}
\begin{abstract}
We report evidence from the 3B Catalogue that long ($T_{90} > 10$ s) and
short ($T_{90} < 10$ s) gamma-ray bursts represent distinct source
populations.  Their spatial distributions are significantly different, with
long bursts having $\langle V/V_{max} \rangle = 0.282 \pm 0.014$ but short
bursts having $\langle V/V_{max} \rangle = 0.385 \pm 0.019$, differing by
$0.103 \pm 0.024$, significant at the $4.3\,\sigma$ level.  Long and short
bursts also differ qualitatively in their spectral behavior, with short
bursts harder in the BATSE (50--300 KeV) band, but long bursts more likely
to be detected at photon energies $> 1$ MeV.  This implies different spatial
origin and physical processes for long and short bursts.  Long bursts may be
explained by accretion-induced collapse.  Short bursts require another
mechanism, for which we suggest neutron star collisions.  These are capable
of producing neutrino bursts as short as a few ms, consistent with the
shortest observed time scales in GRB.  We briefly investigate the parameters
of clusters in which neutron star collisions may occur, and discuss
the nuclear evolution of expelled and accelerated matter.
\end{abstract}



\section{Introduction}

\cite{K93} found that the durations of ``classical'' gamma-ray
bursts\footnote{This paper does not discuss soft gamma repeaters.} (GRB)
are bimodally distributed and are anti-correlated with their spectral
hardness as measured by BATSE in the 50--300 KeV band.  These authors found
no statistically significant differences between the spatial distributions
of long and short GRB.

Little attention has been paid to the cause of the division of GRB into two
classes, long and short.  The simplest interpretation is that all GRB are of
similar origin, but have different values of one or more bimodally
distributed parameters (for example, the interstellar density [\cite{K94a}]
or the shock parameters [\cite{SP95}]).
Most models of GRB have several poorly determined parameters.  The observed
anti-correlation of duration with spectral hardness is naturally obtained in
the neutrino-fireball-debris-shock (NFDS) process (\cite{RM92,MR93,K94a}) in
which a higher Lorentz factor $\Gamma$ leads to a shorter GRB and to a
higher characteristic synchrotron frequency $\nu_{Synch}$.

The most popular NFDS models have been based on coalescing neutron stars
(\cite{E89}).  However, in gravitational radiation-driven coalescence of a
binary star the radial component of relative velocity of the stars is
subsonic with respect to their interior sound speed.  Unless the
viscosity is spectacularly large ($\sim 10^{29}$ gm cm$^{-1}$ s$^{-1}$), the
flow is nearly adiabatic, with little heating or neutrino emission.  The
only strong shock expected would be where a mass-transfer stream strikes an
accreting star or disc, a process which does not occur for stars of similar
mass and radius, as in a neutron star binary.  Three-dimensional
hydrodynamic calculations (\cite{JR96,Ma96}) confirm this conclusion; there
is insufficient heating or neutrino emission to create an energetic fireball.

Accretion-induced collapse (AIC) of a bare degenerate dwarf has been
calculated (\cite{D92}) to produce sufficient neutrino flux to power a
fireball.  AIC produces a $\sim 10$~s neutrino burst, as observed from
SN1987A (the presence of a stellar envelope turns the neutrino energy into a
supernova, while its absence permits a relativistic fireball).  The duration
of neutrino emission is a lower bound on the duration of the resulting GRB
because the subsequent shock interaction, particle acceleration and
radiation can stretch the observed GRB (as will the cosmological redshift),
but cannot so readily shorten it.  We note that AIC is much more frequent
than GRB, which require that the collapse be surrounded by the right density
of matter to produce a relativistic baryonic shell.

AIC can therefore explain only long GRB with $T_{90} > 10$ s; shorter GRB
require a different process.  Both long and short GRB must occur at
``cosmological'' distances (great enough that the geometry of space is
non-Euclidean or cosmological evolution is significant) because each class
is isotropically distributed on the sky and has $\langle V/V_{max} \rangle <
0.5$ (\cite{K93}).  Both probably produce soft ($< 1$ MeV) gamma-rays by the
NFDS process, explaining their qualitatively similar soft gamma-ray spectra
and complex pulse forms, but the origins of the energy must be different.

The purpose of this paper is to investigate the hypothesis that long and
short GRB have different physical origins.  In \S2 we compare their spectra
at soft and harder gamma-ray energies, and find a qualitative difference.
In \S3 we compare their distributions in space, and find a quantitative
difference, although both classes are at ``cosmological'' distances.  These
two independent conclusions each confirm our hypothesis that short and long
GRB are the result of different events.  AIC is discussed as the origin of
long GRB in \S4.  Short GRB require a new process, for which we suggest
colliding neutron stars in \S5.  The nuclear composition of expelled and
accelerated matter, and its implications for GRB, are briefly estimated in
\S6.  \S7 contains a general discussion.

\section{Spectral Behavior}

If long and short GRB are produced by two distinct kinds of events, members
of these two classes may have qualitatively different spectral properties.
This hypothesis can be tested with data in the 3B catalogue (\cite{Me96}).
We find that there are indeed qualitative differences between the spectral
properties of long and short GRB.

The BATSE hardness ratio is a measure of the spectral slope in the range
50--300 KeV.  Some of the bursts in the 3B catalogue were also detected by
COMPTEL, EGRET and OSSE, indicating the presence of energetic photons
above the BATSE band.  The data are summarized in Table \ref{1}.  The choice
of a BATSE hardness ratio criterion of 10 in the first line of the Table
selects the very hardest BATSE spectra, but is necessarily arbitrary.  This
was a natural choice when we first scanned the data by eye, and we have
retained it.

It is evident that long and short GRB differ qualitatively.  Considering
only the BATSE and COMPTEL detections, the probability of this distribution
(or of a greater difference between short and long GRB) being obtained from
a single population of events is $< 10^{-8}$.  The EGRET detections
strengthen this conclusion, while the two OSSE detections contribute little
information.  We cannot, however, exclude the possibility of a systematic
bias in detection thresholds of COMPTEL (in comparison to BATSE) in favor of
detection of long GRB.

Photons with energies $> 1$ MeV are detected almost exclusively from long
($T_{90} > 10$~s) GRB.  This is opposite to the behavior expected from an
extrapolation of the hardness measured at lower photon energies by BATSE.
Long and short GRB show qualitatively different spectral behavior, which
cannot be explained by variation of a single parameter, such as
$\nu_{Synch}$.  We conclude that different physical processes must be
involved, rather than different parameter ranges of a single process.

\section{Spatial Distributions}

If long and short GRB have distinct physical origins they may have distinct
spatial distributions, although this is not specifically predicted.  Table
\ref{2} presents the results of an analysis of the $C/C_{min}$ data in the
3B Catalogue (\cite{Me96}).  The Whole Catalogue analysis shows that $\langle
V/V_{max} \rangle$ for long and short GRB differ by $4.3\,\sigma$, which is
very unlikely to be a statistical fluctuation.  Long and short GRB are
thus found to have different spatial distributions, and therefore must
originate in different classes of events.

We analyzed the positional data in the 3B Catalogue separately for long and
short GRB, and found no statistically significant dipole or quadrupole
deviations from isotropy for either class (confirming the result of
\cite{K93}).  We conclude that both classes are at ``cosmological''
distances, but that long GRB are more deficient in faint bursts, and are
therefore closer, on average.  This presumably reflects differences in the
cosmological evolution of the two source populations.

C. Kouveliotou (private communication) suggested that we subdivide the data
on the basis of integration times, as shown in Table \ref{2}.  This shows
that small $\langle V/V_{max} \rangle$ is chiefly a property of the
subpopulation of smoothly rising GRB (most of which are also long), which
are defined by the criterion that they do not trigger the detector when short
integration times (64 ms) are used.  We can predict that if these ``smooth
risers'' could be separated from other long GRB they would have even smaller
$\langle V/V_{max} \rangle$, and would be a pure AIC population,
uncontaminated by GRB which have an intrinsically short time scale (and
hence rapid rise) but whose $T_{90}$ is long because they radiate slowly or
have multiple widely separated subpulses.  Individual identification of
``smooth risers'' would require the complete time-histories of each GRB, and
would be difficult because of the limited signal-to-noise ratio of most GRB.

\section{Long GRB}

Any model of long GRB must explain their $> 1$ MeV emission by a process
distinct from that which produces their lower energy emission.  This
process must occur only in long GRB, implying that their physical
environment is fundamentally different from that of short GRB.  Such a model
was developed (\cite{K94b}) for the extraordinarily intense burst 3B940217,
which was very long ($T_{90} = 150$ s) as measured by BATSE, and which also
produced photons of energies as high as 18 GeV an hour after the initial burst
(\cite{H94}), but which had the unremarkable BATSE hardness ratio of 3.83.  In
this model the energetic gamma-rays are attributed to $\pi^0$ decay or to
Compton scattering by energetic electrons and positrons (themselves produced
by $\pi^\pm$ decay) resulting from relativistic nuclei (fireball debris)
colliding with baryons in a dense cloud of circumfireball matter.  We now
suggest that such a model is applicable to many or all long GRB (but no
short GRB), though the density and geometry of the cloud will necessarily
vary from event to event, as will therefore the efficiency of production of
energetic gamma-rays.

The cloud was attributed to excretion by the progenitor of one of the
neutron stars in a coalescing neutron star model.  That specific scenario
must be replaced by one of AIC: when matter flows into an accretion disc
surrounding the degenerate dwarf a fraction $f$ of it is excreted from the
disc and the binary.  This is inevitable; matter accreting onto the dwarf
must give up nearly all its angular momentum, which flows outward by viscous
stress in the accretion disc.  Conservation of angular momentum gives
$$f = 1 - ({r_{RCR} / r_{LSO}})^{1/2}, \eqno(1)$$
where $r_{RCR}$ is the Roche circularization radius (\cite{K73}) and
$r_{LSO}$ is the radius of the last stable disc orbit (\cite{B74}), from
which mass peels off the disc and is lost; $f \approx 0.5$, almost
independent of the binary mass ratio.

The circumfireball cloud must be rather small ($< 10^{15}$ cm) in order that
it be dense enough for collisional interaction with the relativistic debris.
For energetic collisional gamma-rays detected simultaneously with a 30 s
GRB the time of flight suggests a size $\sim 10^{12}$ cm, but this may be an
underestimate (by a factor up to $\Gamma^2$) if the relativistic particles
are moving radially outward at the time of collision; as a result, the time
of flight may not give a useful bound on the cloud dimensions.  However, the
requirement of collisional interaction gives a secure lower bound to the
density and therefore, for reasonable cloud masses, an upper bound to the
cloud's dimensions, independent of any assumptions about relativistic
kinematics.

A degenerate dwarf cannot accrete hydrogen-rich matter at a rate faster than
$3 \times 10^{-7} M_\odot$ y$^{-1}$ because the Eddington limit bounds its
thermonuclear luminosity.  As a result, AIC is likely to be preceded by a
period of accretion of order the Eddington time (\cite{K87})
$$t_E = {\epsilon c \kappa \over 4 \pi G} \approx 3 \times 10^6\,{\rm y},
\eqno(2)$$
where $\epsilon c^2$ is the thermonuclear energy release per gram and
$\kappa$ is the opacity.  It is not known how close to the Chandrasekhar
limit is the degenerate dwarf when it begins accretion, but known degenerate
dwarfs are at least a few tenths of $M_\odot$ below that limit, implying
accretion over at least $\sim 10^6$ y.

Even the largest possible cloud is much too small to be freely expanding
over an accretion time of $\sim 10^6$ y.  It could be gravitationally bound
in an excretion disc outside the binary orbit, although it is not known how
long such a disc would survive.  Alternatively, accretion of helium or
carbon-oxygen matter could proceed much faster because of the reduced
thermonuclear energy release, efficient neutrino cooling (in burning of
carbon and heavier elements) and the difficulty of igniting these fuels.
Accretion of heavier elements resembles degenerate dwarf coalescence more
than conventional mass transfer, and might be rapid enough (gravitational
radiation-driven coalescence lasts $\sim 30$ y) that escaping matter would
still be sufficiently close and dense when the final collapse occurred, even
were it freely escaping.

Apart from their gamma-ray emission, such events might roughly resemble
supernovae, as energy deposited in the cloud is thermalized and radiated.
If classed as supernovae, they may be of unusual type and
subtypical luminosity and duration (because the cloud is probably less
massive than typical supernova envelopes).  The predicted gravitational wave
emission of a long GRB, produced by AIC, is $\sim 10^{-9} M_\odot c^2$
(\cite{K80,BH96}), or even less if no matter is expelled.

\section{Short GRB}

Short GRB require a new mechanism.  The requirement of producing $\sim 10^{51}$
erg of soft gamma-rays, and $\sim 10^{53}$ erg of neutrinos if the NFDS
process is assumed, points to a catastrophic event involving one or more
neutron stars; half the energy must be released in 10 ms in at least a few
GRB.  We suggest the collision of two neutron stars, probably occurring in a
very dense cluster of stars.

\subsection{Colliding neutron stars}

Unlike a mass transfer binary, colliding neutron stars will not generally be
surrounded by a massive cloud.  Any such cloud would probably be dispersed when
the neutron stars were born; if not, it would rapidly be disrupted in a
dense star cluster.  Hence collisional production of pions and high energy
gamma-rays is not expected from short GRB, consistent with the rarity of their
detection by COMPTEL (the few detections of short GRB might be of the high
energy tail of the soft gamma-ray emission at $\sim 1$ MeV) and the absence
of their detection by EGRET or OSSE.

Colliding neutron stars move on nearly parabolic orbits before collision.
For masses of $1.4 M_\odot$ and typical equations of state (\cite{WFF88})
they have velocities (with respect to their center of mass) of $\approx
0.62c$ at contact, normally directed in a head-on collision.  This is mildly
supersonic in matter with the typical mean neutron star density $\rho_{ns}
\approx 7 \times 10^{14}$ g cm$^{-3}$, for which these equations of state
yield a sound speed $\approx 0.45c$.  The resulting shock, requiring a
supersonic velocity of
convergence not found in coalescing binary neutron stars, is Nature's way of
making the large dissipation required for a GRB from a small viscosity.

The collision of two neutron stars is a very complex process, involving a
strongly non-ideal equation of state, three-dimensional (unless the
collision is head-on) hydrodynamics, and significant effects of general
relativity.  However, a rough estimate may be useful.  The potential energy
density attributable to the encounter is $GM \rho_{ns}/s$, where $M$ is the
mass of each neutron star and $s$ a mean separation.  The internal energy of
a shock-heated neutron star interior is ${11 \over 4}aT^4$ if only photon
and electron and muon neutrino (and anti-neutrino) specific heats are
considered.  This yields an under-estimate of the internal energy, for it
neglects the contributions of all charged particles, but is unlikely to be
far wrong: the high Fermi energies of neutrons and electrons reduce their
specific heats significantly from their nondegenerate values and limit the
production of electron-positron pairs, protons are scarce, and muon pairs
are massive enough that comparatively few are produced.

If, in addition, we neglect the increase in density upon collision and the
fraction of the energy release which appears as adiabatic compression of the
degenerate matter rather than as thermal energy, we can equate the
available and thermal energy densities:
$${GM\rho_{ns} \over s} = {11 \over 4} a T^4. \eqno(3)$$
For $M = 1.4 M_\odot$, $\rho_{ns} = 7 \times 10^{14}$ g cm$^{-3}$ and $s = 2
\times 10^6$ cm (corresponding to first contact in the absence of tidal
distortion) we find $k_B T \approx 115$ MeV.  Some of our approximations
tend to cancel but most are in the direction of over-estimating $T$.  The
fourth power of $T$ in Eq. 3 is forgiving, so it is probably fair to assume
an initial post-collision temperature $k_B T_0 \approx 100$ MeV.

The post-collision configuration has enough energy to recreate its initial
state of two neutron stars with zero velocity at infinity.  The collision
redistributes energy, so that some matter may escape with speed $v$
at infinity, leaving the remainder bound in a single object (which may
promptly collapse to a black hole).  We take a mass $M_e$ expelled
into a solid angle ${\hat \Omega} \le 4\pi$ sterad,
beginning from a region of size $r_0$ at temperature $T_0$.

At first this expelled matter is opaque to neutrinos because of its high
density and temperature.  We estimate its neutrino diffusion time $t_{diff}
\approx 3 r^2 \rho \sigma/(m_H c)$, where $\sigma \approx 1.7 \times
10^{-38} (k_B T/100{\rm MeV})^2$ cm$^2$ is a mean neutrino interaction
cross-section (\cite{JR96}) and $m_H$ is the nucleon mass.  Adopting $\rho
\approx 3 M_e/({\hat \Omega} r^3)$ and using the adiabatic cooling law for a
relativistic gas of photons and neutrinos $T \propto \rho^{1/3} \propto
r^{-1}$, we equate $t_{diff}$ to the hydrodynamic expansion time $r/v$ to
estimate the radius $r$ at which most of the internal energy is radiated as
neutrinos.  The result is
$$r \approx \left({9 \over {\hat \Omega}}{v \over c}{M_e \over m_H}\,1.7
\times 10^{-26}\right)^{1/4} \left({k_B T_0 \over 100\,{\rm
MeV}}\right)^{1/2} \left({r_0 \over 10^6\,{\rm cm}}\right)^{1/2}\ {\rm cm}.
\eqno(4)$$
For ${\hat \Omega} = 3$ sterad, $v = 10^{10}$ cm s$^{-1}$, $M_e = 0.3
M_\odot$, $k_B T_0 = 100$ MeV and $r_0 = 10^6$ cm we find $r \approx 5
\times 10^7$ cm.  The characteristic width of the neutrino pulse is $\sim
r/(2v) \approx 2.5$ ms.  The escaping neutrinos can then make a relativistic
pair fireball in near-vacuum outside the expelled matter.  The neutrino
pulse width is a lower bound on the duration of the ultimate GRB.  This
result is consistent with all GRB durations and is not far from the shortest
time-scales observed in GRB; both these facts support this model of the
physical processes in short GRB.

A simple estimate shows that the gravitational radiation emitted in the
collision of two neutron stars is $\sim 10^{-2} GM^2/r \sim 10^{51}$
erg into a broad band around $\sim 3$ KHz.  The wave-train would be very
different from that of coalescing neutron stars.

\subsection{Clusters of neutron stars}

A cluster of radius $R$, containing $N$ stars each with mass $M$ and radius
$r_s$, has an evaporation time
$$t_{ev} \approx {200 N \over \ln N} t_{cr}, \eqno(5)$$
where $t_{cr} \equiv (R^3/GMN)^{1/2}$ is the crossing time.  The time scale
for the cluster to evolve by collisions is
$$t_{coll} \approx {R \over r_s} t_{cr}, \eqno(6)$$
where the cross-section, allowing for gravitational focusing and nearly
parabolic orbits as the neutron stars approach each other, is
$$ \sigma \approx r_s R/N \eqno(7)$$
if $R/N \gg r_s$, a condition met in all clusters of interest.  The total
collision rate is
$$\nu_{coll} \sim \left({GM r_s^2 N^3 \over R^5}\right)^{1/2} \sim 10^{19}\,
{N^{3/2} \over R^{5/2}}\,{\rm s}^{-1}. \eqno(8)$$

Allowed parameter regimes are shown in Figure \ref{F1}, in which the stars
have been taken to be neutron stars.  Relativistic
instability is avoided if $\Omega \equiv GMN/(Rc^2) \lesssim 0.1$.  The
present upper bound (\cite{Me95}) on the repetition rate of GRB is $\sim
10^{-8}$ s$^{-1}$.  If $N_S$ short GRB were detected over a period $t_{obs}$
with angular accuracy $\delta \theta \ll (4 \pi)^{1/2}/N_S$ (so that
accidental coincidences are negligible), repetition rates as small as $\sim
(N_S t_{obs})^{-1}$ could be detected.  This could be a much more stringent
bound than that set by BATSE data, whose large positional uncertainties
introduce a substantial background of accidental coincidences.  Note,
however, that an unknown fraction of neutron star collisions produce
observable GRB, so that $\nu_{coll}$ may exceed the observed repetition
rate of short GRB.

It is unclear whether the cluster evolution time should be longer or shorter
than the age of the Universe.  The evolution which produced the cluster
must, of course, require no more than $\sim 10^{10}$ y, but may be
shorter than the evolution time of the GRB-emitting cluster itself (for
example, if the earlier evolution involved collisions of less compact stars
with larger cross-sections).  A long-lived (less dense) cluster may have
produced GRB for most of the age of the Universe, and will do so for a very
long time, but a certain fraction of shorter-lived clusters would be active at
any given time.  A good analogy is to globular clusters, which are observed
with core collapse times both longer and shorter than the age of the
Universe.  As a result, it is impossible to exclude any region of Figure
\ref{F1} except those with $\nu_{coll} \gg 10^{-8}$ s$^{-1}$ or $0.1
\lesssim \Omega$.  Clusters with high collision rates may permit the
observation of repeating GRB, but there is no {\it a priori} reason to
expect this.

A hypothetical cluster with $N = 10^8$ and $R = 10^{18}$ cm (virial velocity
$\approx 2 \times 10^8$ cm/s) has a collision rate $\sim 10^{-14}$
s$^{-1}$ and a lifetime of $\sim 10^{19}$ s.  About $10^9$ such clusters
would be required to produce the observed $10^{-5}$ short GRB s$^{-1}$
within $z \sim 1$; we cannot exclude that such clusters are commonly found at
the centers of galaxies.

Dense clusters involving frequent collisions were discussed (\cite{G65}) as
a possible origin of quasars, but this is now considered unlikely because
stellar collisions do not obviously account for the nonthermal particle
acceleration processes which are the essence of the active galactic nucleus
phenomenon.  (GRB are evidence, however, for particle acceleration in
unexpected circumstances!)  Quasar models make extreme demands on cluster
parameters, for a quasar luminosity of $10^{46}$ erg s$^{-1}$ requires a
collision rate about $10^7$ times that of our hypothetical cluster, if each
collision releases $10^{53}$ erg of observable energy (most of which is
actually lost as neutrinos, widening the disparity).  Cluster models
intended to explain quasars were therefore very dense, and suffered from
short lifetimes or relativistic instability.  The cluster parameters
required to explain GRB as the consequence of neutron star collisions are
much less extreme, making such clusters more plausible.

Our hypothetical cluster would probably be undetectable at ``cosmological''
distances, except for its rare GRB activity.  Its mean collisional
luminosity, including neutrinos, is only $\sim 10^{39}$ erg s$^{-1}$.
Because the collision cross-section and rate scale as the $+1$ power of the
stellar radius (Eq. 7), and the specific binding energy as the $-1$ power,
the collisional luminosity is roughly independent of stellar radius for a
cluster of specified $R$, $N$ and $M$.  A cluster with similar values of
these parameters, but less compact stars, would have a similar collisional
luminosity but more frequent collisions and more rapid collisional evolution.

It is now considered likely that many galaxies possess massive black holes
at their centers, which plausibly grew from dense clusters of stars.  When
the density becomes high collisions become frequent, and lower density stars
are disrupted, leaving only neutron stars and black holes.  Dense clusters
of evolved stars are plausible precursors to, or companions of, massive
black holes; such a black hole has little effect on the structure of the
cluster unless the black hole's mass is dominant, a possibility we ignore
(it increases the stellar velocities, and requires another parameter to
describe it).

\section{Thermonuclear Processing}

In any NFDS model of GRB some material is accelerated from a dilute
fireball above a neutrinosphere to make the relativistic debris shell.  In
AIC this material has very high entropy; a 10 s wind carrying
$10^{-8}\,M_{\odot}$ from the surface of a neutron star has a density there
of $\sim 10$ g cm$^{-3}$, but a temperature $\sim 1$ MeV (\cite{D92}).  A
relativistic debris shell produced by a neutron star collision is formed
under roughly similar fireball conditions, but may have a density $\sim
10^3$ g cm$^{-3}$ because its duration may be $\sim 10^{-3}$ s and its
surface area of origin may be $\sim 10^{15}$ cm$^2$ (\S5).  Expelled neutron
star debris, not accelerated to relativistic velocity, has much lower
entropy, for it is shock heated to $k_B T \sim 100$ MeV at $\rho \sim
10^{14}$ g cm$^{-3}$.  Rather few baryons are accelerated in
ultrarelativistic fireballs, but a much larger mass (perhaps $\sim
0.1\,M_\odot$) of neutron star matter may be expelled at subrelativistic
speed.

The nuclear composition of these several sources of matter is of interest.
Temperatures of 1 MeV at densities $\ll 10^{-8}$ g cm$^{-3}$ are
sufficient to dissociate all nuclei to their constituent neutrons and
protons, as will the much higher temperatures found behind shocks in neutron
star interiors.  The neutrino flux above a neutrinosphere is insufficient to
equilibrate neutron and proton numbers in the time available ($\sim 10^{-4}$
s) in the accelerating flow (\cite{W72}).  In contrast, in a shocked neutron
star interior with a black-body neutrino density at $k_B T \sim 100$ MeV
neutron-proton equilibrium will be achieved rapidly.

In either case, once the matter cools by adiabatic expansion nucleosynthesis
will begin.  The problem resembles that of nucleosynthesis in the early
Universe, but with higher density and shorter time scales.  At $k_B T
\lesssim 100$ KeV equilibrium favors the reaction ${\rm p} + {\rm n} \to {\rm
D} + \gamma$, and unless the density is very low ($< (4 \times 10^4
t_{exp})^{-1}$ g cm$^{-3}$), where $t_{exp}$ is the expansion time, all of
the less numerous (of p and n) species will be bound as deuterons.  Neutron
beta decay is very slow, and the familiar network of reactions among p, n,
D, T and $^3$He, which rapidly convert nearly all the D to $^4$He, follows.
The 3-$\alpha$ reaction is too slow to be significant, so the products
are almost entirely $^4$He and either n or p.  In contrast to the case of
adiabatic decompression of neutron star matter (\cite{E89}), no r-process or
other heavy nuclei are produced.

The neutrons decay on a length scale (in the local observer's frame) $\sim
\Gamma c t_n$, where the neutron decay time $t_n \approx 1000$ s.  For
relativistic debris this may be comparable to the scale of shock interaction
with the surrounding medium.  Neutrons change this interaction.  They run
ahead of the shock itself because they are not slowed by electromagnetic
fields.  Their decay introduces a stream of relativistic protons and
electrons into the medium; this mix of counterstreaming charged particles is
subject to plasma instabilities, and if equipartition is achieved produces
synchrotron radiation analogous to (and comparable to) that calculated in
shock models of GRB.  The subsequent charged particle shock enters a
relativistically pre-heated medium, and is weak.

\section{Discussion}

Our choice of $T_{90} = 10$ s as the dividing line between long and short
GRB was made {\it a priori}, on the basis of the theoretically predicted
duration of the neutrino burst from AIC (empirically supported by
observations of neutrinos from SN1987A).  Our criterion differs from that (2
s) suggested by \cite{K93}.  They chose this value because it is the
observed minimum of the distribution of $T_{90}$.  There are not many GRB
with 2 s $< T_{90} <$ 10~s.  We found that these have $\langle V/V_{max}
\rangle$ indistinguishable from those of even shorter GRB, and that GRB
with 10 s $< T_{90}<$ 20 s are indistinguishable from even longer GRB.  This
supports our choice of a 10 s criterion.  It is important, however, to
define the division into long and short classes before analyzing the data
(as we did), rather than searching for a criterion which maximizes the
effect, which would make the error of using an {\it a posteriori} test of
statistical significance.

The division of GRB into two classes implies that statistical tests (for
isotropy, repetitions {\it etc.}) should be performed on each class
separately.  It is possible that members of each class may have different
properties, even though all may radiate by the NFDS process, and effects
which are statistically significant for one class may not be significant for
the entire population.  For example, we determined the angular
autocorrelation function separately for long and short GRB, in order to
search for repetitions in each class of source without the statistical
interference of random coincidences with members of the other class, but
found no statistically significant effect.  We also searched for dipole and
quadrupole anisotropy in each class, but again found no significant effect.

Each of the models for GRB discussed here probably implies a surrounding gas
density, with which the relativistic shock interacts, much greater than the
typical interstellar density $\sim 1$ cm$^{-3}$ usually assumed.  For long
GRB, a cloud of mass 0.3 $M_\odot$ and radius $10^{15}$ cm has a mean
density $\sim 10^{11}$ cm$^{-3}$, although out of the binary orbital plane
it is likely to be much more dilute.  The gas density within a dense cluster
of neutron stars is difficult to estimate, but is increased above ordinary
interstellar densities by accretion from the surrounding media and by tidal
or collisional disruption of the occasional nondegenerate star which wanders
into the cluster.  The consequences of higher values of the gas density
include shorter GRB pulses and more efficient conversion of the kinetic
energy of relativistic debris to radiation.

\acknowledgments
This research has made use of data obtained through the Compton Gamma-Ray
Observatory Science Support Center Online Service, provided by the
NASA-Goddard Space Flight Center.  We thank J. Clark, R. Kippen and C.
Kouveliotou for discussions and NASA NAGW-2918 and NAG-52862 and NSF
AST 94-16904 for support.

\newpage

\begin{table}
\caption{BATSE hardness and higher energy detections; nominal sensitivity
ranges are indicated.  Data are from the 3B Catalogue.}
\label{1}
\begin{tabular}{lcc}
 &$T_{90} < 10$ s& $10\ {\rm s} < T_{90}$\\
\tableline
BATSE Hardness Ratio $> 10$ (0.05--0.3 MeV) & 21 & 1 \\
COMPTEL Detections (1--30 MeV) & 4 & 20 \\
OSSE Detections (0.06--10 MeV) & 0 & 2 \\
EGRET Detections (20--30,000 MeV) & 0 & 6 \\
\end{tabular}
\end{table}

\begin{table}
\caption{$\langle V/V_{max} \rangle$ for short and long GRB.  Data from 3B
Catalogue.}
\label{2}
\begin{tabular}{lcccc}
 & Whole Catalogue & 64ms Data & 256ms Data & 1024ms Data \\
\tableline
$T_{90} < 10$ s & $0.385 \pm 0.019$ & $0.383 \pm 0.021$ & $0.373 \pm 0.027$
& $0.391 \pm 0.022$ \\
$T_{90} > 10$ s & $0.282 \pm 0.014$ & $0.370 \pm 0.018$ & $0.305 \pm 0.017$
& $0.276 \pm 0.014$ \\
Difference & $0.103 \pm 0.024$ & $0.013 \pm 0.028$ & $0.069 \pm 0.032$
& $0.114 \pm 0.026$ \\
\end{tabular}
\end{table}

\newpage
\figcaption{Parameter regimes for clusters of colliding neutron stars.
\label{F1}}
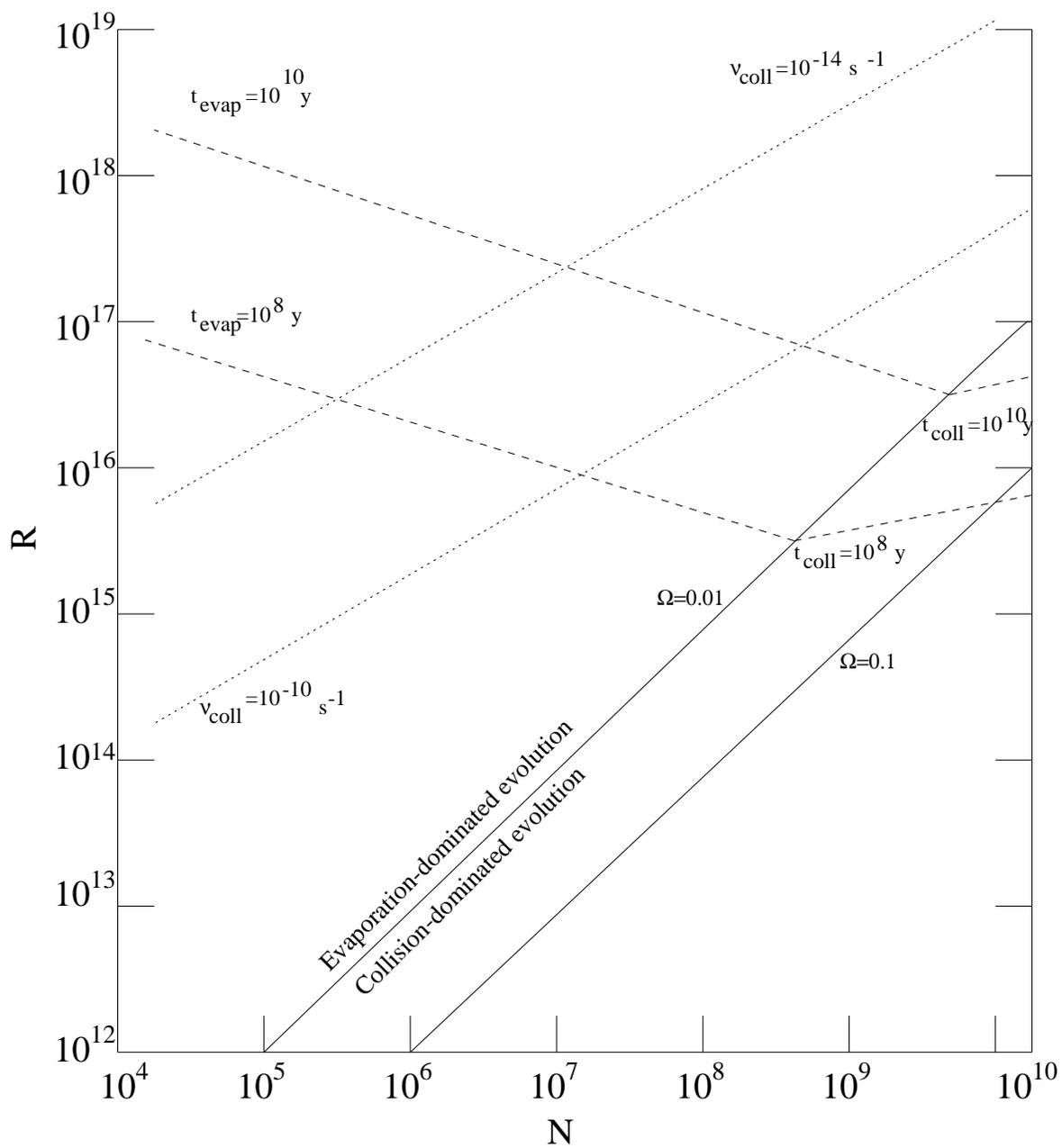
\end{document}